\documentclass[prl,amsmath,amsfonts,amssymb,twocolumn,superscriptaddress,showpacs]{revtex4}

\usepackage{graphicx,psfrag}
\usepackage{dcolumn}
\usepackage{bm}
\usepackage{mathbbol}
\usepackage{makeidx}               
\usepackage{graphicx}              
\usepackage[bookmarksnumbered,colorlinks,plainpages,citecolor=blue,urlcolor=blue]{hyperref}
\usepackage[rflt]{floatflt}
\usepackage{subfigure}
\usepackage{caption2}
\usepackage{latexsym}
\usepackage{amsmath}
\usepackage{amsfonts}
\usepackage{amssymb}
\usepackage{enumerate}

\newcommand{\uck}[1]{\o}

\newcommand{\ket}[1]{\mbox{$|#1\protect\rangle$}}

\newcommand{\bra}[1]{\mbox{$\protect\langle#1|$}}

\def\beq{\begin{equation}}
\def\eeq{\end{equation}}
\def\bea{\begin{eqnarray}}
\def\eea{\end{eqnarray}}

\begin{document}



\title{Realisation of Qudits in Coupled Potential Wells}
\author{Ariel Landau$^1$, Yakir Aharonov$^{1,2}$, Eliahu Cohen$^3$ \\
\small $^1$School of Physics and Astronomy, Tel-Aviv University, Tel-Aviv 6997801, Israel\\
\small $^2$Schmid College of Science, Chapman University, Orange, CA 92866, USA\\
\small $^3$H.H. Wills Physics Laboratory, University of Bristol, Tyndall Avenue, Bristol, BS8 1TL, U.K}



\maketitle

\section{Abstract}Quantum computation strongly relies on the realisation, manipulation and control of qubits. A central method for realizing qubits is by creating a double-well potential system with a significant gap between the first two eigenvalues and the rest.
In this work we first revisit the theoretical grounds underlying the double-well qubit dynamics, then proceed to suggest novel extensions of these principles to a triple-well qutrit with periodic boundary conditions, followed by a general $d$-well analysis of qudits.
These analyses are based on representations of the special unitary groups $SU(d)$ which expose the systems' symmetry and employ them for performing computations. We conclude with a few notes on coherence and scalability of $d$-well systems.

\section{Introduction}

The use of quantum mechanical systems for information processing and computation has been studied extensively over the last decades. Models have been well established for using the principles of quantum mechanics to achieve computational advantages in the form of a speedup over classical methods, and to allow efficient simulation of physical systems. However, a central challenge on the way to fulfilling these goals continues to lie in the realisation of such systems - how to construct a large-scale system, which entails the required quantum properties and at the same time allows efficient manipulations necessary for performing these tasks.

Some of the main techniques for the realisation problem feature the use of a \textit{double-well potential} system to store and manipulate the qubit. Realisations of a double-well-based qubit have been demonstrated using superconducting circuits (\textit{SQUID}), which provide efficient control over the qubit transformations and feasible integration with electronic circuitry \cite{Ultrafast SQUID, Flux-based qubit}.

As an alternative to the qubit, work has been done to study the analogue 3-state register, the \textit{qutrit}, and more generally the $d$-state \textit{qudit}. Several advantages of using qutrits and qudits rather than qubits have been discussed, e.g. in the context of improved fault-tolerance \cite{Fault tolerance qudits, Fault Tolerance 2} and advantages in cryptography \cite{QKD qudits, Qutrit Trapped Ions}.
Motivated by these results we suggest a novel \textit{periodic triple-well qutrit} system. On the one hand, it is based on the known control and manipulation principles of the double-well qubit, while on the other hand it presents special symmetries beneficial to the higher dimensional analysis of its evolution and control using the fundamental representation of $SU(3)$. We further opt for a generalization of these concepts to a general dimension periodic $d$-well qudit, with a central role played by the respective representation of $SU(d)$.

The paper is organized as follows: Sec. 1 reviews the dynamics of a particle in a double-well potential while emphasizing its close relations with a spin-1/2 system. We mostly discuss in this introductory section the methods for control and manipulation of the qubit state with the underlying goal of efficient execution of general transformations. This simple analysis will be hopefully useful in the rest of the paper which manifests its main contribution.
Sec. 2 proposes the extension to a periodic triple-well qutrit system, discussing its dynamics and manipulation with regard to a set of transformations in $SU(3)$.
Sec. 3 aims to further extend the above ideas in a system of $d>3$ wells, and in particular the corresponding evolution of a qudit in the scope of $SU(d)$. Two different topologies are considered for the problem: (i) a fully-connected well system featuring a highly symmetrical solution; (ii) a cyclic linearly-connected well chain, featuring a Bloch-state based solution, and essentially of higher feasibility.

\section{1. Revisiting the Double-Well Potential}
\label{chap:Double-Well}

\subsection{Review of States and Dynamics}

The double-well potential system is a widely considered tool for producing a 2-level system, due to its energy separation featuring a pair of lower states well-separated from the rest of the spectrum:

\begin{equation} \label{Energies_low_states}
\Delta E_{01} \ll \Delta E_{12}.\\
\end{equation}

Eq. \ref{Energies_low_states} can be derived by solving the Schr\"{o}dinger equation for the specific double-well potential form. A solution for the simple square-well case is summarized in \textit{Appendix A} and further detailed in \cite{exact_double_well_schrodinger}.\\

The energy gap $\Delta E_{01}$ is related to the tunneling amplitude, denoted by $\nu$. As denoted for example in  \cite{LandauLifshitz chap 7}, this tunneling amplitude can be related to the wavefunction using the WKB approximation

\begin{equation} \label{tunneling amplitude eq}
\nu =  \frac{\hbar^2}{m} \left( \psi_s \frac{d\psi_s}{dx} \right) \bigg|_{x=0} \ ,
\end{equation}
where $\psi_s$ is the localized wavefunction in the right or left well. The solution can generally be written as \cite{Energy splitting double-well}
\begin{equation} \label{tunneling amplitude eq}
\nu =  C \cdot \exp\left( -\frac{2}{\hbar} \int_{-a/2}^{a/2}dx\sqrt{2m(V(x) - E)} \right),
\end{equation}
for a barrier of width $a$ around the origin\footnote{C can be expressed as a polynomial, expanded in powers of $\hbar$.}. For example, in the square potential (see \textit{Appendix A}) we obtain for $\nu$ \cite{Energy splitting double-well},

\begin{equation} \label{Two symmetric well states}
\nu = \frac{2\hbar E \sqrt{2m(V_0 - E)}}{mV_0 L} \exp\left( -\frac{2a}{\hbar} \sqrt{2m(V_0 - E)} \right).
\end{equation}

The proportion between $\nu$ and $\Delta E_{01}$ can be simply found in a two-state system analysis, performed in the next section.

Solving the time-dependent Schr\"{o}dinger equation, it is apparent that the tunneling implies oscillatory dynamics of the 2-level state. Shall we define $\ket{L}$ and $\ket{R}$ as projections on the left and right wells, and take initial conditions of $\ket{\Psi(t=0)} = \ket{L}$ for example, the right-well probability amplitude $P_R = \langle R \ket{\Psi}$ then oscillates:
\begin{equation} \label{symmetric_dynamics_example}
\frac{dP_R}{dt} = \frac{\nu}{\hbar} \sin(\omega t),
\end{equation}
with a frequency
\begin{equation} \label{Rabi frequency}
\omega = \frac{\Delta E_{01} }{ \hbar}.
\end{equation}

In general we use the following criteria to define a double-well system, in either one or higher dimensions:
\begin{enumerate}
\item Parity symmetry: the two wells are identical under a discrete symmetry denoted by $P$, e.g. reflection about $x=0$ for a 1D well $PV(x) = V(-x)$ or about a center plane in 3D, $PV(x,y,z) = V(x,y,-z)$ \footnote{Alternative 3D generalizations may also apply, e.g. $PV(x,y,z) = V(-x,-y,-z)$.}. The commutation relation is therefore satisfied:
\begin{equation} \label{parity_commutation}
[H,P] = 0.
\end{equation}
\item The coupling between the two wells is weak enough, i.e. the barrier is high enough, so that Eq. \ref{Energies_low_states} applies.
\end{enumerate}

These criteria suffice to produce a symmetric and anti-symmetric pair of lower states, $\ket{0}$ and $\ket{1}$, which serve as a 2-level system in the low-energy regime (with amplitudes for higher levels negligible, $\langle k\ket{\Psi} \approx 0$ ; $k \geq 2$).

\subsection{2-level operators}

Setting a reference basis of \textit{right and left projections}, we may write a general two-state symmetric Hamiltonian for the system:
\begin{equation} \label{qubit_Hamiltonian}
\begin{array} {lcl}H_{2,s} = -\nu\sigma_x = -\nu\begin{pmatrix}
0 & 1 \\
1 & 0
\end{pmatrix},
\end{array}
\end{equation}
with $\nu$ the tunneling amplitude (Eq. \ref{tunneling amplitude eq}), and the basis vectors
\begin{equation} \label{su2_base_def}
\ket{\uparrow} =
\begin{pmatrix}
1 \\
0
\end{pmatrix}, \ket{\downarrow} =
\begin{pmatrix}
0 \\
1
\end{pmatrix}
\end{equation}
are localized states in the right/left well, namely $\ket{R},\ket{L}$.\\

The Hamiltonian's invariance to a $\uparrow \Leftrightarrow \downarrow$ swap manifests the parity symmetry defined in Eq. \ref{parity_commutation}.
Solving for the energy eigenstates relates the tunneling amplitude with the energy gap:
\begin{equation} \label{Symmetric eigen energies}
E_{0,1} = \pm \nu \quad \Rightarrow \quad \Delta E_{01} = 2\nu.
\end{equation}

Thus, remaining in the low energy regime with significant amplitudes held only for the states $\ket{0}, \ket{1}$, we may treat the system as equivalent to a spin-1/2 particle, under the generating group $\frac{1}{2}\sigma_i$ with $\sigma_i$ being the Pauli matrices:
\begin{equation}
\sigma_x=
\begin{pmatrix}
0 & 1 \\
1 & 0
\end{pmatrix}
 ,\ \sigma_y=
\begin{pmatrix}
0 & -i \\
i & 0
\end{pmatrix}
,\ \sigma_z=
\begin{pmatrix}
1 & 0 \\
0 & -1
\end{pmatrix}.
\end{equation}

The eigenvectors of the simplified Hamiltonian $H_{2,s}$ are the $\sigma_x$ eigenstates:
\begin{equation} \label{Two symmetric well states}
\begin{array} {lcl}
\ket{0} =
\frac{1}{\sqrt{2}}
\begin{pmatrix}
1 \\
1
\end{pmatrix}
 =  \ket{+}, \quad
 \ket{1} =
\frac{1}{\sqrt{2}}
\begin{pmatrix}
1 \\
-1
\end{pmatrix}
 = \ket{-}. \\
\end{array}
\end{equation}
Satisfying the $P$ symmetry, the energy states $\ket{0}, \ket{1}$ are \textit{symmetric} and \textit{antisymmetric}, respectively, in context of the $P$ operator.\\

Mapping the $\sigma_z$ and $\sigma_x$ operators to location and energy, respectively, the third generator $\sigma_y$ remains to be addressed. We shall demonstrate that the corresponding physical quantity for this spin operator is the \textit{probability current} between the two wells.\\
The positive/negative probability current values account for the change of total density from the left to the right well and vice versa,
\begin{equation}
\begin{array} {lcl}
J = i(\ket{L}\bra{R} - \ket{R}\bra{L}).
\end{array}
\end{equation}
In our spin states notation, the flow operator hence acts as $\frac{d}{dt}\sigma_z$ in the Heisenberg picture. Using our previous identifications of $\sigma_z \Leftrightarrow \ket{R}\bra{R} - \ket{L}\bra{L}$ and $\sigma_x  \Leftrightarrow H$ we derive in the Heisenberg picture:
\begin{equation}
\begin{array} {lcl}
J = \frac{d}{dt}\sigma_z = \frac{i}{\hbar}[H,\sigma_z] = -i\frac{\nu}{\hbar}[\sigma_x,\sigma_z]  = i\frac{\nu}{\hbar}\sigma_y
\end{array}.
\end{equation}
Hence, the probability current between wells is proportional to the $\sigma_y$ in the 2-level notation, with the coefficient $\frac{\nu}{\hbar}$ corresponding to the maximal current magnitude.\\
The current eigenstates are therefore the $y_+,~y_-$ vectors:
\begin{equation}
\begin{array} {lcl}
 \ket{y_+} =
 \frac{1}{\sqrt{2}}
\begin{pmatrix}
1 \\
i
\end{pmatrix},  \qquad
\ket{y_-} =
 \frac{1}{\sqrt{2}}
\begin{pmatrix}
1 \\
-i
\end{pmatrix}.
\end{array}
\end{equation}
It is straightforward to show that these states also produce maximal (positive) and minimal (negative) values of probability current at $x=0$:
\begin{equation}
\begin{array} {lcl}
\frac{dP_L}{dt} = -\frac{dP_R}{dt} =  \int_{0}^{\infty}\frac{d\rho(x)}{dt}dx = j(x)\vert_{x=0}\ ,
\end{array}
\end{equation}
where $j(x) = \frac{\hbar}{2mi}\left(\Psi^*\frac{d\Psi}{dx}-\Psi\frac{d\Psi^*}{dx} \right)$, while using the continuity equation in the rightmost equality.

The relation of energy, current and position to the qubit operators $\sigma_x$, $\sigma_y$, $\sigma_z$ provides physical meaning to the different qubit transformations, explicitly connecting each one with a corresponding observable.

\subsubsection{Dynamics and manipulation}
\label{sec:Qubit Manipulation}

With the Hamiltonian $H_{2,s}$, the time evolution of the double-well qubit corresponds to precession about the $x$-axis, equivalent to a spin-1/2 particle in a magnetic field $B\hat{x}$.

In general, the evolution of a state $\ket{\psi} = \alpha\ket{0} + \beta\ket{1}$ is:
\begin{equation}
\begin{array} {lcl}
\ket{\psi(t)} = e^{-\frac{i}{\hbar}Ht}\ket{\psi(0)} = e^{-\frac{i}{\hbar}E_0t}\alpha\ket{0} + e^{-\frac{i}{\hbar}E_1t}\beta\ket{1}\\[2mm] =e^{-\frac{i}{\hbar}E_0t}(\alpha\ket{0} + e^{-i\omega t}\beta\ket{1}).\\
\end{array}
\end{equation}
Hence the evolution of a qubit state is periodic:
\begin{equation} \label{qubit periodicity}
\begin{array} {lcl}
\ket{\psi(t+T)} = e^{i\alpha}\ket{\psi(t)},
\end{array}
\end{equation}
with a global phase $\alpha$, and a period $T$ determined by the Rabi frequency:
\begin{equation} \label{Rabi Freq eq}
\begin{array} {lcl}
\omega = \frac{2\pi}{T} = \frac{E_1 - E_0}{\hbar}.
\end{array}
\end{equation}

If we intend to perform a transformation on the double-well qubit, let us first ask what set of qubit transformations would be sufficient. In order to carry out a general unitary transformation, one would alter the Hamiltonian to a modified form
\begin{equation} \label{Hamiltonian as Magnetic Field}
H' = \vec{B}\cdot\vec{\sigma} = B_x\sigma_x + B_y\sigma_y + B_z\sigma_z,
\end{equation}
for a specific time period, generating a precession about an axis defined by the $\vec{B}$ direction (using the notation of a spin-1/2 particle in a magnetic field), its magnitude reflecting the precession frequency.
The original system Hamiltonian (Eq. \ref{qubit_Hamiltonian}) refers to the case $B_x=-\nu, B_y=0, B_z=0$.

As we note later on, to perform a general qubit transformation it is sufficient to control only two of the field components, for instance:
\begin{equation} \label{asymmetric spin hamiltonian}
H' = B_x\sigma_x + B_z\sigma_z.
\end{equation}
The choice of $x$ and $z$ is not incidental; As we show in the following, this form can be realized by inducing asymmetry in the well system.

\subsubsection{Dynamics in an asymmetric well}
\label{sec:Asymmetric}

We examine the effect of the asymmetry on the eigenstates and energy eigenvalues. Using the notation of Rastelli et al. \cite{Asymmetric double-well}, the asymmetry is quantified by a dimensionless parameter $\eta$, i.e. $\eta=0$ in the symmetric case and $\eta\neq0$ otherwise. Our two basis vectors remain the localized states $\Psi_L(x), \Psi_R(x)$ with \textit{local} energies $\epsilon_L, \epsilon_R$, respectively, found by solving the Schr\"{o}dinger equation locally in the left and right wells. \footnote{This is done by neglecting the tunneling, assuming the low energy regime.}. The difference between the energies $\Delta \epsilon = \epsilon_L - \epsilon_R$ is related to the difference between the well depths. \footnote{In case the wells are of identical shape, the differences are equal: $\Delta \epsilon = \min V_L(x) - \min V_R(x)$ (denoting by $L$,$R$ the $x<0$, $x>0$ sides, respectively, and taking the local minima as benchmarks).}

\begin{figure}[h]
 \centering \includegraphics[height=5cm]{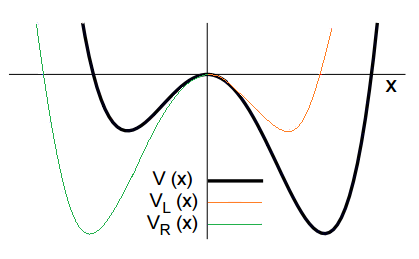}
      \caption{{\bf Construction of the symmetric potentials $V_L(x,\eta), V_R(x,\eta)$.}} \label{Asymmetric_well_reflections}
\end{figure}

As derived by Rastelli et al. in \cite{Asymmetric double-well}, the value of $\nu(\eta)$ in the presence of asymmetry can be derived using the WKB approximation to be

\begin{equation} \label{Asymmetric_tunneling_amplitude1}
\nu(\eta) = A(\eta) \sqrt{\nu_L(\eta) \nu_R(\eta) } ,
\end{equation}
where
\begin{equation} \label{Asymmetric_tunneling_amplitude2}
A(\eta) = \frac{1}{2} \left[ \left( \frac{V(0)-\epsilon_L(\eta)}{V(0)-\epsilon_R(\eta)} \right)^{\frac{1}{4}} +  \left( \frac{V(0)-\epsilon_R(\eta)}{V(0)-\epsilon_L(\eta)} \right)^{\frac{1}{4}}\right],
\end{equation}

 and $\nu_L(\eta), \nu_R(\eta)$ are obtained by two \textit{symmetric} double-well potentials $V_L(x,\eta)$ and $V_R(x,\eta)$, defined by a symmetric reflection of the left or right side of the asymmetric well, as demonstrated in Fig. \ref{Asymmetric_well_reflections}.

Using the basis states $\Psi_{L/R}(x)$ in the regime of Eq. \ref{Energies_low_states}, the ground state and first excited state are \\
\begin{equation}
\begin{split}
\Psi_0(x) = \sin\left(\frac{\theta}{2}\right)\Psi_L(x) + \cos\left(\frac{\theta}{2}\right)\Psi_R(x), \\
\Psi_1(x) = \cos\left(\frac{\theta}{2}\right)\Psi_L(x) - \sin\left(\frac{\theta}{2}\right)\Psi_R(x),
\end{split}
\end{equation} \\
with the angle $\theta$ related to the Hamiltonian parameters by:
\begin{equation} \label{Angle and Params relation}
\theta = \arctan \left( \frac{\nu(\eta)}{\Delta\epsilon(\eta)/2} \right).
\end{equation}


To what degree of asymmetry is this result valid? If the potential is tilted to a certain level, the local ground state of one well begins to interact with the excited state of the other, lower well. Thus, our results hold as long as the energy difference $\Delta \epsilon$ is not comparable to the one-side energy gap between ground and first excited state,
\begin{equation} \label{Regime for theta}
\Delta \epsilon < \epsilon_{s,1} - \epsilon_{s,0}\ ,
\end{equation}
where $s$ stands for the side of low potential. \\

Moving on to examine the effects in the 2-level system, the modified Hamiltonian depends on the degree of asymmetry $\eta$, and reflects the different energies of the $\ket{L},\ket{R}$ states with a tunneling amplitude $\nu(\eta)$:
\begin{equation} \label{Perturbed Hamiltonian}
H'(\eta) = \begin{pmatrix}
\epsilon_R(\eta) & -\nu(\eta) \\
-\nu(\eta) & \epsilon_L(\eta)
\end{pmatrix}.
\end{equation}

Removing a $\frac{1}{2}(\epsilon_L+\epsilon_R) I$ factor as a physically irrelevant global phase, we are left with the Hamiltonian
\begin{equation} \label{Perturbed Hamiltonian 2}
H'(\eta) = \frac{1}{2}\Delta\epsilon(\eta) \sigma_z - \nu(\eta) \sigma_x,
\end{equation}
where $\Delta\epsilon = \epsilon_L-\epsilon_R$.

The implication of this asymmetry on the direction of $\vec{B}$ in the Hamiltonian (Eq. \ref{Hamiltonian as Magnetic Field}), is simply setting $\vec{B}$ in an angle $\theta$ (Eq. \ref{Angle and Params relation}) from the $\hat{z}$ direction in the $XZ$ plane.

In the case of symmetric energies, $\epsilon_L = \epsilon_R$, the resulting angle is $\theta = \frac{\pi}{2}$ and $\vec{B}$ is aligned with $\sigma_x$, resorting to the symmetric and antisymmetric states of the symmetric double-well in Eq. \ref{Two symmetric well states}.

The two-state eigenvectors of the asymmetric system are thus

\begin{equation} \label{Two asymmetric well states}
\begin{array} {lcl}
\ket{0(\theta)} =
\begin{pmatrix}
\sin\frac{\theta}{2} \\[1mm]
\cos\frac{\theta}{2}
\end{pmatrix} , \quad
 \ket{1(\theta)} =
\begin{pmatrix}
\cos\frac{\theta}{2} \\[1mm]
-\sin\frac{\theta}{2}
\end{pmatrix}.
\end{array}
\end{equation}

The two energy eigenvalues are straightforwardly found from Eq. \ref{Perturbed Hamiltonian 2} to be
\begin{equation} \label{Asymmetric energies}
\Delta E_{01} = 2\sqrt{ \left( \frac{\Delta\epsilon}{2} \right) ^2 + \nu^2(\eta) },
\end{equation}

reducing to Eq. \ref{Symmetric eigen energies} for $\Delta \epsilon = 0$.

The asymmetric eigenstates $\ket{0(\theta)}, \ket{1(\theta)}$ hence imply a modified dynamics of the qubit state. The state now `oscillates' around $\vec{B}(\theta)$, rather than $\sigma_x$. The Rabi frequency $\omega = \frac{\Delta E_{01} }{ \hbar}$ is now increased relative to the symmetric case of Eq. \ref{Symmetric eigen energies}.

\subsection{Implementation: The controllable SQUID double-well}

An instructive example for a double-well implementation, extensively studied and tested, is via the use of superconductor devices, based on the Josephson effect \cite{Josephson qubits}, and specifically, the RF-SQUID circuit \cite{Ultrafast SQUID, Wendin}. A full analysis of this device, namely the \textit{double SQUID}, is presented in \cite{Ultrafast SQUID}. We shall briefly describe the system and introduce its resulting Hamiltonian.

\begin{figure}[h]
 \centering \includegraphics[height=4cm]{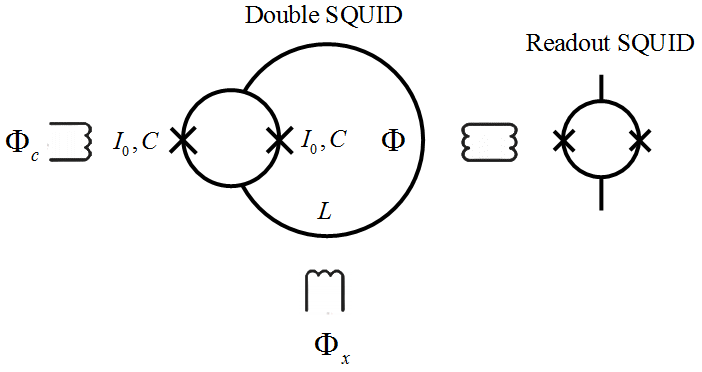}
      \caption{\bf The double SQUID setup with readout coupling.} \label{Double_SQUID}
\end{figure}
 A scheme of the double SQUID is given in Fig. \ref{Double_SQUID}. The two sides of the junction with capacitance $C$ and critical current $I_0$ are connected by a superconducting loop with inductance $L$\footnote{The loop is normally a few square microns in size\cite{Flux-based qubit}.} (the inner loop inductance is assumed negligible). An external flux $\Phi_{x}$ is imposed through the loop by an auxiliary coil.
 Taking the flux through the superconducting loop $\phi=\frac{\Phi}{\Phi_b}$ as the quantum variable ($\Phi_b = \frac{\Phi_0}{2\pi}$ with $\Phi_0$ the flux quantum, $\Phi_0 = \frac{h}{2e} = 2.07\cdot10^{-15}Wb$), and its conjugate variable the charge on the capacitance, $p$, the resulting Hamiltonian is
\begin{equation} \label{SQUID_Hamiltonian}
H = \frac{p^2}{2M} + \frac{\phi_b^2}{L}\left[ \frac{1}{2} (\phi - \phi_{x})^2 - \beta(\phi_c)\cos(\phi) \right],
\end{equation}\\
where $p=Q\phi_b$ is the conjugate momentum of $\phi$, with $Q$ the charge on the capacitor; $M=C\Phi_b^2$ is the effective mass; $\beta(\phi_c) = \left(\frac{2LI_0}{\phi_b}\right)$ ;  $\phi_x$ and $\phi_c$ are the control fluxes in reduced units $\phi_x = \frac{\Phi_x}{\Phi_b}, \phi_c = \frac{\Phi_c}{\Phi_b}$.

\begin{figure}[h]
 \centering \includegraphics[height=4.2cm]{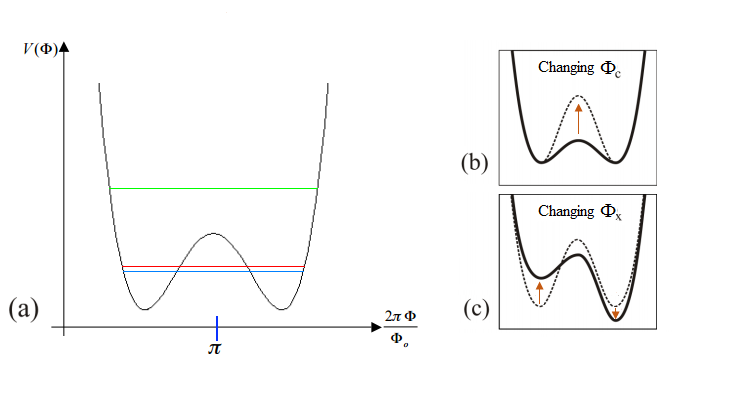}
      \caption{{\bf Double SQUID potential plots.} (a) A symmetric double-well example, where the energies of the ground state (blue), first excited state (red) and second excited state (green) are denoted. (b) The effect of the control flux $\Phi_c$ on the potential barrier. (c) The effect of the control flux $\Phi_x$ on the symmetry. } \label{SQUID_Well}
\end{figure}

The resulting potential is a continuous double-well as shown in Fig. \ref{SQUID_Well}a. The potential barrier height is modified by changes in $\Phi_c$ (\ref{SQUID_Well}b). An asymmetry in the wells' energies may be induced by tuning $\Phi_x$ (\ref{SQUID_Well}c). The resulting two-state system is also referred to as a \textit{flux qubit}. Note that the state readout is done via coupling of the flux with a second SQUID, activated when the measurement is needed. However, other readout methods exist. \cite{Ultrafast SQUID, Andreev readout}

Note that in practice, to maintain quantum coherence in a given two-state system, it has to be cooled down to very low temperatures. For the system to be robust to thermal excitations, the energy gap has to significantly exceed the thermal energy:

\begin{equation} \label{Thermal fluct}
k_b T \ll \Delta E_{01}.
\end{equation}

This condition sets a lower bound to the Rabi frequency in the high barrier state of $\Phi_c$. A temperature of $1K$ corresponds to a frequency of $20GHz$, while the typical range of the SQUID qubit is around $1\sim20GHz$ (see \ref{sec:Qubit Manipulation}); thus it is required to cool the system to about $20mK$ or below \cite{SCqubits_Review}.\\
Other noise sources contribute to the decoherence of the qubit state, which vary in dominance per system conditions. These factors normally divide into two classes - energy relaxation (drifting towards the lower energy state) and dephasing (diffusing of the vector in the longitudinal direction). For further details and analysis of decoherence factors, see \cite{SCqubits_Review, flux noise}. 

We also note that other realisations have been made for the double-well potential, e.g. double quantum dots \cite{quantum dots} or interacting Bose-Einstein condensates trap \cite{be trap}. Nevertheless, the superconducting realisation is of central influence, considering its well-established design in a scalable size and possibility of on-chip integration with electronic circuits \cite{Ultrafast SQUID,Flux-based qubit}.

\subsection{Qubit operations}

We may now use the above derivations to describe a procedure of double-well qubit transformation: A desired operation on the `spin' is translated to a sequence of changes of the potential features - barrier height and asymmetry - which in turn translate in the double SQUID to appropriate pulses in the fluxes $\Phi_c, \Phi_x$ \cite{Ultrafast SQUID, IBM qubit 2006}.
The common method of control is via RF-pulses throughout the coils; however, in their work, Castelli et al.  introduce a short pulse method  for `ultra-fast' manipulation of the qubit \cite{Ultrafast SQUID}. Further implementation related questions of operation frequency and coherence time are addressed later in this section.

We list and show the qubit operations in the system; First rotations about the X- and Z- axes, then a tilted XZ axis, and finally the generation of rotations about any axis in the Bloch sphere using the former steps.

\subsubsection{X-rotation}

Rotation about the $x$-axis corresponds to $B_z = 0$ in the Hamiltonian (\ref{Perturbed Hamiltonian 2}), or simply the unperturbed Hamiltonian (\ref{qubit_Hamiltonian}). The Rabi oscillation frequency may be changed by adjusting the tunneling amplitude ($\Phi_c$ control). In \cite{Ultrafast SQUID}, a rapid lowering of the barrier was used for reaching a single-well condition, before resetting the high barrier after some time $\Delta t$. The rotation angle $\alpha$ then depends on the precession time by $\alpha \approx \omega \Delta t$. The potential change cannot be too fast though, or it would excite upper energy levels\footnote{A typical range for the energy gaps may be around $\Delta E_{21} \approx 20\Delta E_{01}$. For details see \cite{nature flux qubit coupled, Ultrafast SQUID}.}. Thus the Rabi frequency varies and the rotation angle is
\begin{equation} \label{rabi rotation angle}
\alpha = \int_{0}^{\Delta t} \omega(t) dt,
\end{equation}
setting $t=0$ as the gate initialization moment.

The resulting qubit gate is an X-rotation, $R_x(\alpha)$:
\begin{equation} \label{Qubit X rotation}
\begin{array} {lcl}
R_x(\alpha)  =
\begin{pmatrix}
\cos \frac{\alpha}{2} && -i\sin \frac{\alpha}{2} \\[1mm]
-i\sin \frac{\alpha}{2} && \cos \frac{\alpha}{2}
\end{pmatrix}.
\end{array}
\end{equation}

For the special case of  $\alpha = \pi + 2\pi n, n\in\mathbb{N}$ we get the NOT gate. 

\subsubsection{Z-rotation}

A rotation about the $z$-axis, or a \textit{phase gate}, can be realized by inducing an amplitude-modulated RF pulse to $\Phi_x$ in the resonance frequency $\omega$. The addition to the Hamiltonian is then
\begin{equation}
\Delta H = \epsilon_{rf}\cos(\omega t) \sigma_z.
\end{equation}
In the rotating frame or reference, the RF-pulse produces a $Z$-rotation with a frequency proportional to $\epsilon_{rf}$. The area of the pulse envelope determines the angle of rotation. As the pulse ends, the state returns to its $x$-precession, and the phase gate is complete as the rotating frame syncs with the original axes \cite{Science z rotation}.

The shaping of the pulse is an important issue. The operations are in general desired to be fast, in order to perform as many operations as possible within the qubit coherence time. On the other hand, if the pulse is too fast, it will excite higher energy levels which we intend to avoid.\\
A solution to this problem was offered by McDermott et al. \cite{pulse train} by using SFQ (Single Flux Quantum) technique: the control flux is changed not in a single pulse but in a series of short and weak pulses, with pulse-to-pulse spacing equal to the period of the oscillation, i.e.
\begin{equation}
\Phi_x(t) = \Phi_0[\delta(t) + \delta(t-T) + ... + \delta(t-(n-1)T)],
\end{equation}
where $\Phi_0 = \frac{h}{2e}$ is the flux quantum\footnote{The quanta $\Phi_0$ equals an integral of the pulse shape; the shape itself is not significant since the time of a single pulse is short enough relative to $T$.}, $T$ is the oscillation period and $n$ is the number of pulses. This approach is analogous to pumping up a swing by giving a short pulse once per cycle rather than forcing the swing throughout its entire movement. Since the pulses add up coherently, the deposited energy scales as $n^2$ and an operation can be carried out in the order of $\sim40$ pulses, with a typical total time of several $ns$ \cite{pulse train}.


\subsubsection{Tilted XZ-rotation}

\begin{figure}[h]
 \centering \includegraphics[height=5cm]{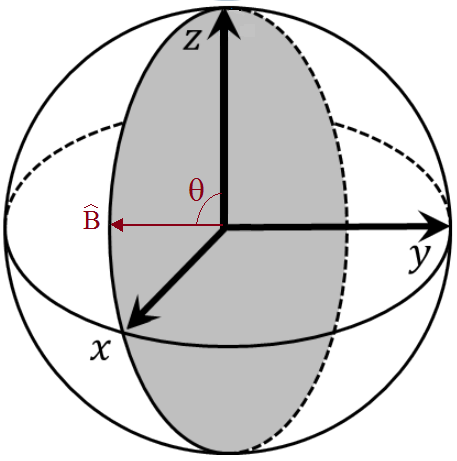}
      \caption{{\bf The $\hat{B}$ axis in the Bloch sphere.}} \label{Bloch_sphere_XZ_plane}
\end{figure}

Using the asymmetry control ($\Phi_x$) we may tilt the rotation axis by an angle $\theta$ (\ref{Angle and Params relation}) from the $z$-axis, and get the general form of $H'$ where $B_x,B_z \neq 0$. The rotation axis $\hat{B} = \frac{\vec{B}}{|\vec{B}|}$ is in the $XZ$ plane of the Bloch sphere, as illustrated in Fig. \ref{Bloch_sphere_XZ_plane}.

This Hamiltonian generates a Rabi oscillation about the tilted $\vec{B}$ axis, in an angle $\alpha$ determined again by Eq. \ref{rabi rotation angle}. However, we note that similarly to the RF-pulse case, the transition of the axis from its previous position is carried out in a finite time in order to not excite high energy levels. The rotation of the state during the time of transition may not be neglected. This issue may be resolved again by using a pulse train approach: The Hamiltonian is changed not in a single pulse but in a series of short and weak pulses. Their spacing is synchronized in each step with the current precession frequency $\omega = \frac{\Delta E_{01}}{\hbar}$ so that for step $i$:
\begin{equation}
T_i = \frac{2\pi}{\omega_i},
\end{equation}
and the pulse train has the form
\begin{equation}
\Phi_x(t) = \Phi_0[\delta(t) + \delta(t-T_1) + ... + \delta(t-\sum_{i=1}^N T_i)].
\end{equation}
This process ends with the $\vec{B}$ axis tilted in the desired angle, without disorienting the state from its initial alignment.

Taking $\vec{n} = (\sin \theta, 0, \cos \theta)$ as the $\vec{B}$ direction, we may write the gate matrix:
\begin{equation} \label{Qubit tilted rotation}
\begin{array} {lcl}
R_{\theta}(\alpha)  = \cos \frac{\alpha}{2} I - i \sin\frac{\alpha}{2} (\sigma_x\sin\theta + \sigma_z \cos\theta ) \\[3mm]
= \begin{pmatrix}
\cos \frac{\alpha}{2} + i\sin\frac{\alpha}{2} \cos\theta && -i\sin \frac{\alpha}{2} \sin \theta \\[1mm]
-i\sin \frac{\alpha}{2} \sin \theta && \cos \frac{\alpha}{2} - i\sin\frac{\alpha}{2} \cos\theta
\end{pmatrix}.
\end{array}
\end{equation}

The special case $\theta = \frac{\pi}{2}$ coincides with (\ref{Qubit X rotation}) as we would expect.

As an example, shall we set the parameters to $2\nu = \Delta \epsilon$ for an angle $\theta = \frac{\pi}{4}$, and a $\Delta t$ which gives $\alpha = \pi + 2\pi k$ (where $k\in \mathbb{N}$), we receive the \textit{Hadamard gate}, namely
\begin{equation} \label{Hadamard Gate}
\begin{array} {lcl}
H = \frac{1}{\sqrt{2}}\begin{pmatrix}
1 && 1 \\[1mm]
1 && -1
\end{pmatrix}.
\end{array}
\end{equation}

How far can $\hat{B}$ be tilted from $\hat{x}$? Eq. \ref{Angle and Params relation} holds in the regime of Eq. \ref{Regime for theta}. For a given $\Delta \epsilon$ within this regime, we may raise the barrier to make $\nu$ small to the extent in which thermal noise $k_b T$ is comparable to the energy gap and decoheres the state. Thus the angle from $\hat{z}$ is roughly lower-bounded by
\begin{equation}
\delta \sim \frac{k_b T}{(\epsilon_{s,1} - \epsilon_{s,0})/2},
\end{equation}
and $\hat{B}$ may be practically set in the range $\theta \in (\delta,\pi-\delta)$. This allows precessions close to the $\hat{z}$ axis but not exactly aligned with it. For an exact $Z$-rotation, the RF method above is available instead.

\subsubsection{The general qubit gate}

Any single-qubit gate can be associated with a certain rotation in the Bloch sphere about an axis $\hat{n}$, which we may denote as:
\begin{equation} \label{n spherical}
\hat{n} = (\sin \theta \cos \psi,  \sin \theta \sin \psi, \cos\theta ).
\end{equation}

This rotation can be achieved in 5 steps of $X$- and $Z$-rotations as follows:
\begin{equation}
R_{\hat{n}}(\alpha) = R_z(\psi')R_x(\theta)R_z(\alpha)R_x(-\theta)R_z(-\psi'),
\end{equation}
with $\psi' = \psi - \frac{\pi}{2}$. The first steps $R_x(-\theta),R_z(-\psi')$ transform $\hat{n}$ to the $\hat{z}$ axis, $R_z(\alpha)$ performs the desired rotation, and finally $R_z(\psi'),R_x(\theta)$ rotate $\hat{n}$ back to its original orientation.\\
By applying the symmetric point $X$-rotation and RF pulse $Z$-rotations described above, any general transformation of the qubit may thus be tailored.

Although this approach is quite intuitive, it is not the most efficient in the number of required steps.
It can be shown \cite{Gates using two axes} that a rotation about a general axis in the sphere, $R_{\hat{n}}(\phi)$, may be performed by a two-step rotation about two axes $\hat{n}_1, \hat{n}_2$ set in one plane, e.g. the $XZ$ plane:
\begin{equation} \label{axes of rotation}
\begin{split}
\hat{n}_1 = (\sin \theta_1, 0, \cos\theta_1 ),\\ \hat{n}_2 = (\sin \theta_2, 0, \cos\theta_2 ).
\end{split}
\end{equation}
 These axes may be set within one half of the plane without loss of generality, i.e. with angles $\theta_1, \theta_2 \in [0,\pi]$.

It is thus straightforward to use the above result for execution of a general single-qubit transformation, using two tilted rotations about proper axes in the $XZ$ plane. We may thus write a general rotation in the Bloch sphere as
\begin{equation} \label{Qubit general rotation}
R_{\hat{n}}(\phi) = e^{i\eta} R_{\theta_2}(\phi_2) R_{\theta_1}(\phi_1),
\end{equation}
where $\eta$ stands for a global phase.

The rotation $R_{\hat{n}}(\phi)$ is characterized by three parameters - the spherical coordinates $\theta, \psi$ of $\hat{n}$ and the rotation angle $\phi$. The right-hand-side however involves five parameters - $\eta, \theta_1, \phi_1, \theta_2, \phi_2$. The equation is therefore under-constrained, and many solutions exist for the rotations $R_{\theta_1}(\phi_1), R_{\theta_2}(\phi_2)$.
Solutions may be analytically constructed in a simple procedure - see \cite{Gates using two axes} for details and examples.

It is therefore shown that the scheme of inducing asymmetry to carry out $XZ$ rotations can be used for a general single-qubit operation in only 2 steps. As far as we know, this operational method has not been proposed so far, whereas the asymmetry was mainly used for preparation in $\ket{L}$ or $\ket{R}$, while followed by the $X$ and $Z$ operations described above. The time scale for this SFQ train would in fact be shorter than in the RF pulse case since the $T_i$ get smaller with $i$; Eventually the arbitrary gate can thus be carried out in the time scale of few $ns$.

\subsubsection{Qubit preparation}

The ability to prepare the qubit state is essential for performing any experiment or computation. In general, preparation of the state can be performed by simply waiting for the system to relax in its ground state.

For instance, preparation of the qubit in a $\ket{\uparrow}$ or $\ket{\downarrow}$ state (left/right well) is done by creating an extreme potential tilt, with a $\Phi_x$ pulse, to the left or right, so that only one well is allowed, and letting the system relax there. This prepares a localization of the system in one side, and following computational operations (changes of $\Phi_c, \Phi_x$) would continue from this point. As mentioned above, this preparation scheme is currently the common use of the $\Phi_x$ bias flux, rather than $XZ$ axis-tilting.

A preparation of the symmetric $\ket{+}$ state can be made similarly by relaxing the system with no $\Phi_x$ tilt, i.e. in the symmetric double-well configuration, where the ground state $\ket{0}$ coincides with $\ket{+}$.\\
For a general initialization along a $\cos\alpha \hat{x} + \sin\alpha \hat{z}$ axis, an appropriate $\Phi_x$ bias is applied for the state to relax there.

\subsubsection{Experimental implementations}

The most promising implementation of a double-well qubit nowadays, as discussed above, is using superconductors and the Josephson effect. Several experiments have realized such systems in attempt to observe high coherence times (limited by unwanted couplings in the system), while maintaining control of the qubit and connectivity in a circuit. By coupling the qubit to a stabilizing harmonic oscillator IBM's system (2006) reached coherence times of 25-35$ns$ \cite{IBM qubit 2006}. Castellano et al. (2010) have demonstrated oscillations of the flux qubit for several cases of lowering the barrier ($\Phi_c$), thus controlling the oscillation frequency in a range of ~10-25$GHz$ with coherence times of order 10$ns$ \cite{Ultrafast SQUID}.

In recent years (2011 - 2013) several groups have begun using a technique of embedding qubits in 3D cavities. This method  has managed to bring the coherence time scale up to 20 - 100 $\mu s$ \cite{Coherence QED cavity 2011, Coherence waveguide cavity 2012}. The use of such qubits in scalable integrated circuits was demonstrated in 2013 by Barends et al. \cite{3D qubit in circuits}. These recent experiments therefore show promising implementations for the double-well Josephson qubit, reaching increasing achievements in the balance of coherence time, control and connectivity \cite{ECBT} \footnote{The quantum computing company D-Wave Inc. have claimed in 2011 to develop a quantum processor demonstrating ad-hoc \textit{quantum annealing} solutions, using a programmable 128 qubit network, rising to 512 qubits in 2012 and 1000+ in 2015 \cite{DWAVE,ECBT2}. The device's quantum qualities, including its true qubit coherence times and a possible quantum speedup have been subject to prolonged debate to this day.}.
It is apparent that progress in the coherence times has an exponential fashion, resembling that of Moore's Law \cite{Qubit Moore Law,Devoret2013}.

The oscillation frequency $\frac{\omega}{2\pi}$ differs among systems in the approximate range of $1\sim 20 GHz$, or periods of $50ps\sim 1ns$\cite{SCqubits_Review}. A gate such as $Z$-rotation is claimed to be carried out as fast as $1-2ns$ by Huang et al. \cite{Optimal control}. Roughly estimating $5ns$ as a sufficient upper limit for arbitrary transformations, recent coherence times of $>1 \mu s$ already fit hundreds, if not thousands, of consecutive single-qubit gates.

The above analysis describes implementation and control of a 2-level register, or qubit, using a double-well. In the following section a generalization of these principles is presented to a three-level system in a triple-well.



\section{2. Up a notch: Triple-Well and $SU(3)$ Computations}
\label{sec:triple well qutrit}


In a manner resembling the double-well case, we present a realisation of a qutrit system using a particle in a spatial potential. For this case, we henceforth define and use the one-dimensional \textit{triple-well} potential, with periodic boundary conditions. The qutrit analysis will be based on the $SU(3)$ group similarly to \cite{QutSU3}.

\subsection{The periodic triple-well system}

The periodic triple-well potential consists of a line (with total length $L$) of three cavities, separated by narrow potential barriers (the identification $x = x+L$ applies). An example for such a physical system would be a particle confined within a ring; taking the angle $\theta$ as the position coordinate, we recognize $\theta \Leftrightarrow x, 2\pi \Leftrightarrow L$, with a potential $V(\theta)$.

\begin{figure}[h]
 \centering \includegraphics[height=5cm]{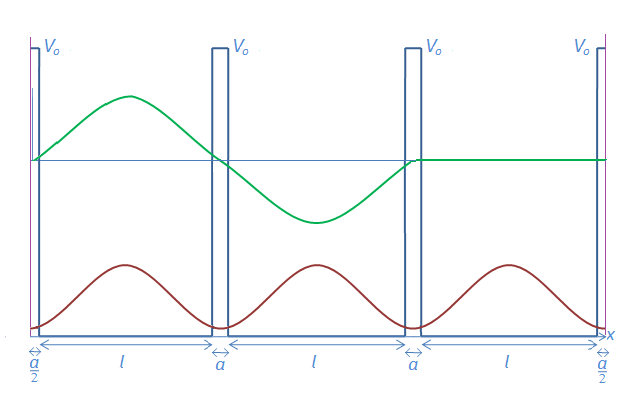}
      \caption{{\bf Periodic triple-well potential in a compact space.} Edge points are identified: $x = x+L $; the ground state (red) and a state in the first excited subspace (green) are illustrated.} \label{Triple_Well_Potential}
\end{figure}

For simplicity, we again illustrate our discussion using a square potential model shown in Fig. \ref{Triple_Well_Potential}.

Nevertheless, this analysis applies for a more general potential as well, should it fulfill the following set of requirements: \\
(i) Translational symmetry: The three wells are equally shaped, namely: $V(x) = V(x+\frac{L}{3})$. In other words, the potential remains unchanged under the translation operation $T_{\frac{L}{3}}(x) \equiv e^{-i\frac{L}{3}px}$. \\
(ii) Parity symmetry: The wells and barriers are symmetric about their center: $V(x_0-x) = V(x_0+x)$, where $x_0$ is the center of a well or a barrier. Given (i), it is then sufficient to require symmetry under the parity operation: $V(x) = V(-x)$ given that $x=0$ is the center of a well.\\
(iii) The barriers are high enough compared to the first energy eigenstates, so that energy separations among the three lowest eigenstates are much smaller than the distance from higher levels:
\begin{equation}
\Delta E_{01}, \Delta E_{12} \ll \Delta E_{23}.
\end{equation}
Similarly to the double-well case, a state of the triple-well system would in effect consist of a combination of three one-well ground states.

Although the square triple-well is not realistic, much like the square double-well, realisations of a smooth form are implementable. For example, a coherent triple-well system has been formed using superconductor circuits analogous to the RF-SQUID, namely a \textit{superconducting qutrit} as described in \cite{SQUTRID}, although it so far lacks the full symmetry required in our discussion. An alternative implementation may be considered using quantum dots for a controllable charge qudit \cite{d-well dots}, as further discussed later in \ref{sec:dwells}, via fabrication of the dots in a triangular topology. In whichever implementation is used, controllability of the potential features is indeed a crucial component, as it sets the ground for executing qutrit transformations (similar to the qubit manipulation discussion in \ref{sec:Qubit Manipulation}).

The eigenfunctions and evolution of the three-well system in the low energy regime may be obtained by the Schr\"{o}dinger equation, in a similar manner to those of the double-well. However, we shall utilize the mathematical simplicity of the 3-state formalism and perform the derivations directly within its frame, in analogy to the two-state analysis of \ref{chap:Double-Well}.

\subsection{Triple-well qutrit states and dynamics}

We first define basis vectors for our $SU(3)$ representation to be the \textit{localized} states of the wavefunction in each of the three wells - similarly to the double-well case. Denoting individual wells by $w_0, w_1, w_2$, our basis vectors are hence the localized states:
\begin{equation}
\begin{array}  {lcl}
\ket{w_0} = \begin{pmatrix} 1 \\ 0 \\ 0 \end{pmatrix}, \ \
\ket{w_1} = \begin{pmatrix} 0 \\ 1 \\ 0 \end{pmatrix}, \ \
\ket{w_2} = \begin{pmatrix} 0 \\ 0 \\ 1 \end{pmatrix}.
\end{array}
\end{equation}

In analogy to the $\sigma_i$ matrices in the 2-level case, we shall make use of the 8-generator set in the fundamental representation of $SU(3)$, the $Gell-Mann$ matrices: $T_a=\frac{1}{2}\lambda_a$, where:
\begin{equation}
\begin{array} {lcl}
\lambda_1=
\begin{pmatrix}
0 & 1 & 0 \\
1 & 0 & 0 \\
0 & 0 & 0
\end{pmatrix}
 ,\ \lambda_2=
\begin{pmatrix}
0 & -i & 0 \\
i & 0 & 0 \\
0 & 0 & 0
\end{pmatrix}, \\[7mm]
\ \lambda_3=
\begin{pmatrix}
1 & 0 & 0 \\
0 & -1 & 0 \\
0 & 0 & 0
\end{pmatrix},
\lambda_4=
\begin{pmatrix}
0 & 0 & 1 \\
0 & 0 & 0 \\
1 & 0 & 0
\end{pmatrix}, \\[7mm]
\ \lambda_5=
\begin{pmatrix}
0 & 0 & -i \\
0 & 0 & 0 \\
i & 0 & 0
\end{pmatrix}
,\ \lambda_6=
\begin{pmatrix}
0 & 0 & 0 \\
0 & 0 & 1 \\
0 & 1 & 0
\end{pmatrix} , \\[7mm]
\lambda_7=
\begin{pmatrix}
0 & 0 & 0 \\
0 & 0 & -i \\
0 & i & 0
\end{pmatrix}
,\ \lambda_8= \frac{1}{\sqrt{3}}
\begin{pmatrix}
1 & 0 & 0 \\
0 & 1 & 0 \\
0 & 0 & -2
\end{pmatrix}.
\end{array}
\end{equation}

These Lie algebra matrices obey the commutation relations
\begin{equation}
\begin{array} {lcl}
[\lambda_i,\lambda_j] = if_{ijk}\lambda_k
\end{array}
\end{equation}
with structure constants $f_{ijk}$, completely antisymmetric in their three indices:
\begin{equation}
\begin{array} {lcl}
f_{123}=1, \\
f_{147}=-f_{157}=f_{246}=f_{257}=f_{345}=-f_{367}=\frac{1}{2}, \\
f_{458}=f_{678}=\frac{\sqrt{3}}{2}.
\end{array}
\end{equation}

The Hamiltonian matrix is constructed by combining the three generators corresponding to two-well transitions:
\begin{equation} \label{eq:H_direction}
\begin{array}  {lcl}
H = -\nu(\sigma^{(01)}_x + \sigma^{(02)}_x + \sigma^{(12)}_x),
\end{array}
\end{equation}
where $\sigma^{(ij)}_x = \ket{w_i}\bra{w_j} + \ket{w_j}\bra{w_i}$.

In the Gell-Mann notation this is equivalent to:
\begin{equation} \label{eq:H_direction}
\begin{array}  {lcl}
H = -\nu(\lambda_1 + \lambda_4 + \lambda_6).
\end{array}
\end{equation}
The Hamiltonian matrix is hence
\begin{equation}
\begin{array}  {lcl}
H = -\nu
\begin{pmatrix}
0 & 1 & 1 \\
1 & 0 & 1 \\
1 & 1 & 0
\end{pmatrix},
\end{array}
\end{equation}
where again, $\nu$ corresponds to the tunneling amplitudes.
It can be easily shown that this is the most general matrix to comply with the system's symmetries.

Using the matrix Hamiltonian we may derive the qutrit eigenvalues and corresponding eigenstates.
Two eigenvalues emerge: \\
The lower energy
\begin{equation}
\begin{array}  {lcl}
E_0=-2\nu,
\end{array}
\end{equation}
corresponds to a \textit{symmetric} eigenvector:
\begin{equation} \label{qutrit_ground_state}
\begin{array}  {lcl}
\ket{0} = \frac{1}{\sqrt{3}}\begin{pmatrix} 1 \\ 1 \\ 1  \end{pmatrix},
\end{array}
\end{equation}
demonstrated by the lower wavefunction in Fig. \ref{Triple_Well_Potential}. \\
The energy of the excited level
\begin{equation}
\begin{array}  {lcl}
E_1=\nu
\end{array}
\end{equation}
corresponds to a \textit{degenerate subspace} of `antisymmetric' states satisfying:
\begin{equation}
\begin{array}  {lcl}
\ket{1} \sim \begin{pmatrix} x \\ y \\ z  \end{pmatrix} ; x+y+z = 0,
\end{array}
\end{equation}
for $x,~y,~z$ being the (complex) amplitudes of $\ket{w_0},~\ket{w_1},~\ket{w_2}$.\\\\
For example, a basis for the $E_1$ subspace may be comprised by the states:
\begin{equation} \label{eq:example_eigenstates_of_E1}
\begin{array}  {lcl}
\ket{1_a} = \frac{1}{\sqrt{2}}\begin{pmatrix} 1 \\ -1 \\ 0  \end{pmatrix}, \ \
\ket{1_b} = \frac{1}{\sqrt{2}}\begin{pmatrix} 0 \\ 1 \\ -1  \end{pmatrix}.
\end{array}
\end{equation}
The upper wavefunction illustrated in Fig. \ref{Triple_Well_Potential} exhibits the form of $\ket{1c_a}$, or equally $\ket{1_b}$ shifted one cell to the left.

Thus a general state of the qutrit may be expanded in terms of these three states:
\begin{equation}
\begin{array}  {lcl}
\ket{\psi} = \alpha\ket{0} + \beta\ket{1_a} + \gamma\ket{1_b}.
\end{array}
\end{equation}
The time evolution of the general state will be described as:
\begin{equation}
\begin{array} {lcl}
\ket{\psi(t)} = e^{-\frac{i}{\hbar}Ht}\ket{\psi(0)} =\\ e^{-\frac{i}{\hbar}E_0t}\alpha\ket{0} + e^{-\frac{i}{\hbar}E_1t}(\beta\ket{1_a} + \gamma\ket{1_b}) = \\
e^{-\frac{i}{\hbar}E_0t}(\alpha\ket{0} + e^{-\frac{i}{\hbar}(E_1-E_0)t}(\beta\ket{1_a} + \gamma\ket{1_b})).
\end{array}
\end{equation}
A key result thus emerges for the symmetric triple-well qutrit system: Any state evolves \textit{periodically} in time, as in the double-well case - even when the Hilbert space is three-dimensional; i.e. the wavefunction undergoes periodic \textit{revival}.\\\\
The periodicity in Eq. \ref{qubit periodicity} applies for the qutrit evolution as well. The revival frequency is
\begin{equation}
\omega = \frac{E_1-E_0}{\hbar} = \frac{3\nu}{\hbar}.
\end{equation}

In analogy with the double-well qubit, The evolution of a qutrit state may be viewed as precession about the Hamiltonian axis (Eq. \ref{eq:H_direction}) in the $SU(3)$ `space' with frequency $\omega$.

Continuing the double-well analogy, let us define the \textit{cyclic current} operator:
\begin{equation} \label{cyclic_current_operator}
J_c =
\begin{pmatrix}
0 & -i & i \\
i & 0 & -i \\
-i & i & 0
\end{pmatrix} = \lambda_2 + \lambda_7 - \lambda_5.
\end{equation}
Unlike the current operator $\sigma_y$ in the qubit case, the cyclic current commutes with $H$. The eigenstates of $J_c$, namely the \textit{current states}, are listed by the corresponding current eigenvalues:
\begin{equation} \label{current_states_triple}
\begin{array} {lcl}
C_0 = 0 ; \;\;\;\;\;\;\;\;\;\;\;\;\;\;\;  \ket{0} = \frac{1}{\sqrt{3}}\begin{pmatrix} 1 \\ 1 \\ 1 \end{pmatrix}\\\\
C_+ = \sqrt{3} ;\;\;\;\;\;\;\;\;\;\;\; \ket{J_+} = \frac{1}{\sqrt{3}}\begin{pmatrix} 1 \\ e^{\frac{2\pi}{3}} \\ e^{-\frac{2\pi}{3}} \end{pmatrix}\\\\
C_- = -\sqrt{3} ;\;\;\;\;\;\;\;\; \ket{J_-} = \frac{1}{\sqrt{3}}\begin{pmatrix} 1 \\ e^{-\frac{2\pi}{3}} \\ e^{\frac{2\pi}{3}} \end{pmatrix}.
\end{array}
\end{equation}\\
A perturbation $H' = H + \epsilon J_c$ would split the $E_1$ degeneracy to different energies for the current states $\ket{J_+},~\ket{J_-}$. The three energy levels after the perturbation are:
\begin{equation}
\begin{array} {lcl}
E_0 = -2h\\
E_{1,+} = h + \sqrt{3}\epsilon,\\
E_{1,-} = h - \sqrt{3}\epsilon,
\end{array}
\end{equation}\\
i.e. the current states evolve differently in a manner determined by the strength of perturbation $\epsilon$.

\subsection{Manipulation}

We have noted that `rotation' of the qutrit state is produced by the 8 generator matrices $T_a=\frac{1}{2}\lambda_a$.

By generalizing the controllable double-well, we shall assume in this triple-well system we have control over:
\begin{enumerate}
\item The height of each barrier, and thus the tunneling between each well couple, $\nu_{ij}$ with the well indices $i,j \in \{0,1,2\}$.
\item The energy difference ('asymmetry') between the adjacent two wells $\Delta \epsilon_{ij}$. The three differences are subject to the periodicity constraint
\begin{equation}
\Delta \epsilon_{01} + \Delta \epsilon_{02} + \Delta \epsilon_{12} = 0.
\end{equation}
\end{enumerate}

A general transformation on a qutrit takes 8 parameters. At a given instant, the control fluxes may be set to control 5 degrees of freedom - the tunneling amplitudes $\nu_{ij}$ and energy differences $\Delta \epsilon_{ij}$. This corresponds to control of the amplitude of 5 generators: the three off-diagonals
\begin{equation}
\ \nu_{01} \lambda_1 \ , \ \ \nu_{12} \lambda_6\ ,\ \ \nu_{01} \lambda_4 ,
\end{equation}
and two diagonals
\begin{equation}
\frac{1}{2}\Delta\epsilon_{01} \lambda_3\ , \ \ \frac{1}{2}\Delta \epsilon_{12} \left( \frac{\sqrt{3}}{2}\lambda_8 - \frac{1}{2}\lambda_3 \right).
\end{equation}

Had we had freedom to shape the control flux pulses as we liked - i.e. instant and high pulses - the control of these components could have allowed any $SU(3)$ transformation we desire: a transformation $U(\vec{\alpha}) \in SU(3)$ may be decomposed into three $SU(2)$ transformations, operating on each pair of indices at a time  \cite{Synthesis of arbitrary $SU(3)$ transformations of atomic qutrits}:
\begin{equation} \label{Qutrit rotations}
U(\vec{\alpha}) = R^{(01)}_{\vec{n}_1}(\phi_1) R^{(02)}_{\vec{n}_2}(\phi_2) R^{(12)}_{\vec{n}_3}(\phi_3),
\end{equation}
with a proper choice of the right-hand-side parameters $\vec{n}_1, \vec{n}_2, \vec{n}_3, \phi_1, \phi_2, \phi_3$. This decomposition may be achieved by fast operation on one pair at a time, by suppressing the other tunneling amplitudes (high barrier) so that the rotation outside the `active' pair is negligible. A transformation $U(\vec{\alpha})$ could be made in as little as 6 steps - two for each double-well rotation, as noted in \ref{sec:Qubit Manipulation}.

However, as we know by now, the strong instant pulses necessary for this scheme cause unwanted excitations and thus do not fulfill the purpose. Changing the fluxes to a certain bias point is spread over a time interval, in which the original state would undergo an unwanted transformation.

A complex ad-hoc calculation may be done for a certain transformation to be carried out via the gradual change of fluxes. Nonetheless, we propose a method to perform a certain subset of transformations on the system.

A general perturbation $\Delta H$ to the Hamiltonian can be written as
\begin{equation}
\begin{array} {lcl}
\Delta H = \Sigma \epsilon_i\lambda_i,
\end{array}
\end{equation}
for $i = 1,...,8$. Thus, perturbing the Hamiltonian towards $H' = H+\Delta H$, sets the time evolution to be:
\begin{equation}
\begin{array} {lcl}
\ket{\psi(t)} = e^{-\frac{i}{\hbar}H't}\ket{\psi(0)} = e^{-\frac{i}{\hbar}(H+\Delta H)t}\ket{\psi(0)}.
\end{array}
\end{equation}

An interesting subset of the possible $\Delta H$ is all the perturbations that \textit{commute with H}, namely $[H,\Delta H] = 0$. In this case the evolution turns out to be:
\begin{equation}
\begin{array} {lcl}
\ket{\psi(t)} = e^{-\frac{i}{\hbar}\Delta Ht}e^{-\frac{i}{\hbar}Ht}\ket{\psi(0)}.
\end{array}
\end{equation}
By applying the perturbation $\Delta H$ for a time $T = \frac{1}{\omega}$, we find
\begin{equation} \label{eq:commuting_perturbation}
\begin{array} {lcl}
\ket{\psi(T)} = e^{-\frac{i}{\hbar}\Delta HT}e^{-\frac{i}{\hbar}HT}\ket{\psi(0)} = \\
e^{i\alpha}e^{-\frac{i}{\hbar}\Delta HT}\ket{\psi(0)},
\end{array}
\end{equation}
namely, that the state is effectively transformed only by the perturbation term:
\begin{equation} \label{eq:perturbation_operator}
\begin{array} {lcl}
U_I = e^{-\frac{i}{\hbar}\Delta HT}.
\end{array}
\end{equation}

Any such operation is part of a subspace of rotations that \textit{commute with H} in $SU(3)$.\\
The basis for this subspace consists of the following four matrices:
\begin{equation}
\begin{array} {lcl}
M_1=
\begin{pmatrix}
-\frac{1}{3} & 1 & 0 \\
1 & -\frac{1}{3} & 0 \\
0 & 0 & \frac{2}{3}
\end{pmatrix},
M_2=
\begin{pmatrix}
-\frac{1}{3} & 0 & 1 \\
0 & \frac{2}{3} & 0 \\
1 & 0 & -\frac{1}{3}
\end{pmatrix}, \\\\
M_3=
\begin{pmatrix}
\frac{2}{3} & 0 & 0 \\
0 & -\frac{1}{3} & 1 \\
0 & 1 & -\frac{1}{3}
\end{pmatrix},
M_4=
\begin{pmatrix}
0 & -i & i \\
i & 0 & -i \\
-i & i & 0
\end{pmatrix}
\end{array}
\end{equation}
or spanned by the Gell-Mann matrices:
\begin{equation}
\begin{array} {lcl}
M_1 = \frac{1}{\sqrt{3}}\lambda_3 + \lambda_1,\\
M_2 = -\frac{1}{2}\lambda_3 + \frac{1}{2\sqrt{3}}\lambda_8 + \lambda_4,\\
M_3 = \frac{1}{2}\lambda_3 + \frac{1}{2\sqrt{3}}\lambda_8 + \lambda_6,\\
M_4 = \lambda_2 + \lambda_7 - \lambda_5.
\end{array}
\end{equation}
Therefore, any transformation generator given by $\epsilon_i M_i$ will commute with $H$, and thus satisfy Eq. \ref{eq:commuting_perturbation}.\\
We divide the above generators into two groups:

(i) The antisymmetric generator $M_4$ is simply the cyclic current operator $M_4 \equiv J_c$ presented above.

(ii) The symmetric generators $M_1,M_2,M_3$:\\
This trio corresponds to certain actions of the control fluxes, i.e. changes of $\nu_{ij}$ and $\Delta \epsilon_{ij}$. Hence, these matrices define a basis for transformations we can induce on the qutrit, based on Eq. \ref{eq:perturbation_operator}.

To execute such a transformation we activate the flux transition in a low magnitude pulses of $\delta H = \epsilon \, \Sigma a_iM_i$, e.g. $\delta H = \epsilon M_1$, over several cycles of $T$ as required for the transformation. Consequently, we may gradually lower it back to $\Delta H = 0$ to restore the revival of $H$. The cumulative energy of the pulses and the time span determine the degree of rotation generated by $\Delta H$. \\ \\
Applying any such perturbation splits the state degeneracy of $E_1$, defining three (generally) non-degenerate eigenstates for $H'$. Taking $\Delta H = \epsilon M_1$ for instance, the emerging eigenstates, apart from $\ket{0}$, are the $\ket{1_a},\ket{1_b} $ defined in Eq. \ref{eq:example_eigenstates_of_E1} above. The corresponding energy levels are:
\begin{equation}
\begin{array} {lcl}
E_0 = -2h + \frac{2}{3}\epsilon,\\[1mm]
E_{1,a} = h - \frac{4}{3}\epsilon,\\[1mm]
E_{1,b} = h + \frac{2}{3}\epsilon.
\end{array}
\end{equation}
Note that $M_2$ and $M_3$ produce similar eigenvectors with respective permutation of their values.
To simplify the action of these generators we transform into a non-traceless form, moving into $U(3)$:
\begin{equation}
\begin{array} {lcl}
M'_i = M_i + \frac{1}{3}I,
\end{array}
\end{equation}
effectively adding a global phase factor to the transformation (i.e. adding a constant $\frac{\epsilon}{3}$ to the energy). We receive the following matrices for $M'_i$:
\begin{equation}
\begin{array} {lcl}
M'_1=
\begin{pmatrix}
0 & 1 & 0 \\
1 & 0 & 0 \\
0 & 0 & 1
\end{pmatrix},\
M'_2=
\begin{pmatrix}
0 & 0 & 1 \\
0 & 1 & 0 \\
1 & 0 & 0
\end{pmatrix}, \\\\ \;\;\;\;\;\;\;\;\;\;\;\;\;\;\;
M'_3=
\begin{pmatrix}
1 & 0 & 0 \\
0 & 0 & 1 \\
0 & 1 &0
\end{pmatrix},
\end{array}
\end{equation}
with unaltered eigenstates and constantly raised energies
\begin{equation}
\begin{array} {lcl}
E_0 = -2h + \epsilon,\\[1mm]
E_{1,a} = h - \epsilon,\\[1mm]
E_{1,b} = h + \epsilon.
\end{array}
\end{equation}
We may calculate the qutrit operator (Eq. \ref{eq:perturbation_operator}) with $\Delta H = \epsilon M'_i$ for each of the $M'_i$:
\begin{equation}
\begin{array} {lcl}
U_1=
\begin{pmatrix}
cos(\epsilon T) & -isin(\epsilon T) & 0 \\
-isin(\epsilon T) & cos(\epsilon T) & 0 \\
0 & 0 & e^{-i\epsilon T}
\end{pmatrix}, \\\\
U_2=
\begin{pmatrix}
cos(\epsilon T) & 0 & -isin(\epsilon T) \\
0 & e^{-i\epsilon T} & 0 \\
-isin(\epsilon T) & 0 & cos(\epsilon T)
\end{pmatrix}, \\\\
U_3=
\begin{pmatrix}
e^{-i\epsilon T} & 0 & 0 \\
0 & cos(\epsilon T) & -isin(\epsilon T) \\
0 & -isin(\epsilon T) & cos(\epsilon T)
\end{pmatrix}.\\
\end{array}
\end{equation}

Trying to interpret the above operations we conclude that each $M'_i$ generates a $\sigma_x$ rotation for the corresponding pair of wells, with a simultaneous $\sigma_z$ rotation adding a relative phase to the amplitude of the third well. The operation angle is given by $\theta = \epsilon T$.\\
We note a specific choice of setting $\epsilon, T$ to produce $\theta = \frac{\pi}{2}$. This results in an action of the $M'_i$ generators directly on the qutrit state, namely, the elementary $X$-gate qutrit operations \cite{Khan, Qutrit computing}:
\begin{equation}
\begin{array} {lcl}
X^{(01)}=
\begin{pmatrix}
0 & 1 & 0 \\
1 & 0 & 0 \\
0 & 0 & 1
\end{pmatrix},\
X^{(02)}=
\begin{pmatrix}
0 & 0 & 1 \\
0 & 1 & 0 \\
1 & 0 & 0
\end{pmatrix}, \\\\ \;\;\;\;\;\;\;\;\;\;\;\;\;\;\;
X^{(12)}=
\begin{pmatrix}
1 & 0 & 0 \\
0 & 0 & 1 \\
0 & 1 &0
\end{pmatrix}.
\end{array}
\end{equation}
The set of $X$-gates provides, for instance, implementation of a set of \textit{ternary shift gates} (see \cite{Khan, Qutrit computing} for further details).

\subsection{Charge measurement and quantum Fourier transform}

Another method is proposed to perform a certain important qutrit gate - the \textit{quantum Fourier transform} (QFT):
\begin{equation} \label{Discrete Fourier Transform 3 level}
QFT_{3\times 3} =
\begin{pmatrix}
1 & 1 & 1 \\
1 & e^{2\pi i/3} & e^{-2\pi i/3} \\
1 & e^{-2\pi i/3} & e^{2\pi i/3}
\end{pmatrix}.
\end{equation}
In a sense, it is a generalization of the Hadamard gate (Eq. \ref{Hadamard Gate}) for more than 2 levels.
The quantum Fourier transform is of main interest in quantum computing, being a key step in Shor's factoring algorithm. The advantageous use of $QFT_{3\times 3}$ to carry out the quantum Fourier transform has been discussed \cite{Qutrit Trapped Ions, ternary QFT}.

The three-state Fourier transformation (Eq. \ref{Discrete Fourier Transform 3 level}) may be achieved by a \textit{change of basis} to the conjugate variable of the flux - the charge in the junctions' capacitance:
\begin{equation}
Q = -i\hbar \frac{\partial}{\partial \Phi}.
\end{equation}
 As conjugate variables, the eigenstates of the charge are Fourier combinations of the flux eigenstates. In our three-state regime, the eigenvalues of the charge operator $\hat{Q}$ are
\begin{equation}
q_- = -2e, \ q_0 = 0, \ q_+ = 2e.
\end{equation}
(note that $2e = \frac{h}{\Phi_0}$).

Therefore, measuring the charge $\hat{Q}$ corresponds to applying the DFT matrix on the qutrit and performing a measurement on the result. Note that the flux states which produce a certain charge eigenvalue with certainty are the eigenstates of the current operator $J_c$.\\

Other methods may be considered to utilize the charge-flux relation for a computation, i.e. coupling of the charge to other qutrits in the system.


\section{3. Generalizing to $d$-Wells}
\label{sec:dwells}

In this section we turn to generalize the three-well qutrit notion to the case of $d$ cyclically coupled wells,
acting as a $d$-level register, i.e. a \textit{qudit}.

It turns out that two different directions may be considered as a canonical generalization of the double- and triple-well. In what follows we present both and then focus on the second which appears to be more practical.

\subsection{The fully-connected $d$-well system}

One proposed generalization is a fully connected network of wells, with equal tunneling amplitudes among each of the pairs. This corresponds to the following $d\times d$ Hamiltonian:

\begin{equation}
\begin{array}  {lcl}
H = -\nu
\begin{pmatrix}
0 & 1 & 1 & \cdots & 1 \\
1 & 0 & 1 &  & 1 \\
1 & 1 & 0 &  &  \\
\vdots & & & \ddots & 1 \\
1 & 1 & & 1 & 0
\end{pmatrix}.
\end{array}
\end{equation}

The system is symmetric under permutations of the wells, and thus is highly symmetric in $SU(d)$. Furthermore, a dynamical analysis reveals but two energy levels: $E_0 = -\nu(d-1)$ for the symmetric ground state,
\begin{equation}
\ket{w_0} = \frac{1}{\sqrt{d}}\begin{pmatrix} 1 \\ 1 \\ \vdots \\ 1 \end{pmatrix}, \
\end{equation}
and the ($d-1$) - degenerate $E_1 = \nu$ corresponding to the orthogonal subspace, i.e. $v_1 + \dots + v_d = 0$, where $v_1,\text{...} \, ,v_d$ are the eigenvector components. As a system of purely two energies, a \textit{state revival} mechanism occurs, much like in the previously discussed qutrit case.

However well-behaved the above generalization is, we also note that a physical realisation of this system may be a challenge, whereas an equal tunneling between all pairs should me maintained. We thus proceed to discuss a different generalization, of a seemingly more feasible nature.

\subsection{The periodic $d$-well system}

\begin{figure}[h]
 \centering \includegraphics[height=5cm]{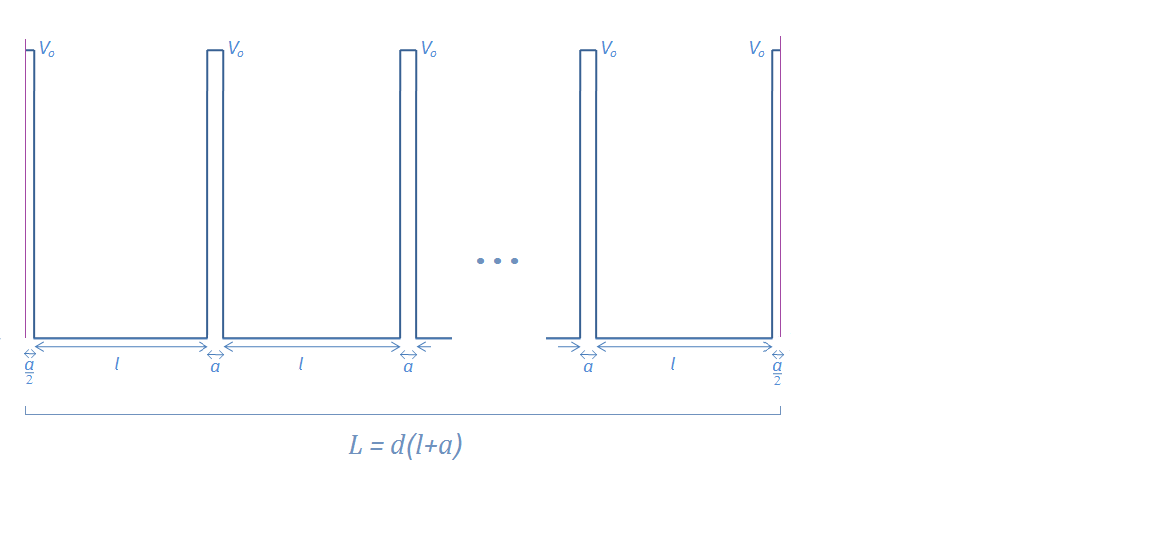}
      \caption{{\bf Periodic $d$-well potential in a compact space.} Edge points are identified: $x = x+L $.} \label{d_Well_Potential}
\end{figure}

A cyclic $d$-well system consists of linearly connected $d$ potential wells, each coupled to its two adjacent wells, once again defined in a compact space, $x = x+L$. The system is illustrated in Fig. \ref{d_Well_Potential}.
The system satisfies the analogue of conditions (i) - (iii) defined above for the three-well system, with the appropriate changes for applying to the $d$ wells case:
\begin{itemize}
    \item The system is symmetric under translations by $\frac{L}{d}$, or: $V(x) = V(x+\frac{L}{d})$..
    \item The resulting spectrum consists of $d$ energy states with separations much smaller than the rest of the states:
    \begin{equation} \label{ddifferences}
    E_{i}-E_{i-1} \ll E_{d}-E_{d-1},
    \end{equation}
    for $i \in \{1, ..., d-1\}$, as shown in the next subsection to result of the Bloch theorem.
\end{itemize}

As in the previous sections, based on these conditions we turn to the $d$-level formalism to derive our results, rather than using the explicit spatial eigenstates.

\subsection{$d$-well qudit states and their dynamics}

The $d$-dimensional basis of the qudit Hilbert space is defined, in a straightforward generalization of the $d=2,3$ cases, as the states of maximally localized wavefunction in each of the $d$ wells, or:
\begin{equation}
\begin{array}  {lcl}
\ket{w_0} = \begin{pmatrix} 1 \\ 0 \\ \vdots \\ 0 \end{pmatrix}, \ \
\ket{w_1} = \begin{pmatrix} 0 \\ 1 \\ \vdots \\ 0 \end{pmatrix}, \ \cdots \ , \
\ket{w_{d-1}} = \begin{pmatrix} 0 \\ 0 \\ \vdots \\ 1 \end{pmatrix}.
\end{array}
\end{equation}

Transformations of the qudit state are matrices in the fundamental representation of the $SU(d)$ group. The generator set consists of $d^2-1$ matrices, spanning all the possible qudit transformations. The symmetric Hamiltonian is of the form

\begin{equation}
\begin{array}  {lcl}
H = -\nu
\begin{pmatrix}
0 & 1 & 0 & \cdots & 1 \\
1 & 0 & 1 &  & 0 \\
0 & 1 & 0 &  &  \\
\vdots & & & \ddots & 1 \\
1 & 0 & & 1 & 0
\end{pmatrix},
\end{array}
\end{equation}
with an identical tunneling amplitude $\nu$ for all neighbouring transitions.

Solving for the system eigenstates, we note that the resulting vectors are \textit{Bloch states}; this is evident from the symmetry and neighbour interactions of $H$, which corresponds to a periodic lattice in a tight-binding regime \cite{SolidState}. The states are eigenstates of the particle's \textit{modular momentum} \cite{Aharonov70}, with its $d$ corresponding eigenvalues
\begin{equation}
\begin{array}  {lcl}
p_{mod,n} = \frac{2\pi n \hbar}{L} \ \ ; \ \ n \in \{0, ..., d-1\}.
\end{array}
\end{equation}

The corresponding energies are given by
\begin{equation} \label{energiesEn}
\begin{array}  {lcl}
E_n = -2 \nu \cos \left(\frac{2\pi (l+a) n}{L}\right).
\end{array}
\end{equation}
We note that apart from the ground state $\ket{p_{mod} = 0}$, and the $
\ket{p_{mod} = \pi \hbar/L}$ state in case of odd $d$ (where $n=d/2$), the spectrum is divided to \textit{degenerate pairs} $\{\ket{p_{mod,n}}, \ket{p_{mod,d-n}}\}$, in agreement with the symmetry of $\cos(x)$ about $x=\pi$. We also note that Eq. \ref{ddifferences} is justified by the model's band structure, featuring a gap between the bands with order of magnitude of the one-well system energy separation $\epsilon_{01}$, much higher than the in-band separations, which are of order $\nu$.


The manipulation of the $d$-well qudit may be similarly performed via control of each potential barrier and each adjacent energy gap. We do not go into a detailed discussion on qudit evolution and operations, as it would follow similar guidelines of the previous $d=2,3$ discussion. We note, however, the following points:
\begin{itemize}
    \item In general, the qudit state does \textit{not} undergo periodic revival, as occurred in the double- or triple-well cases. This is due to the fact that more than two energy values exist among the state basis vectors. Nevertheless, certain cases do feature periodic revival, such as $d=4$ and $d=6$, where the energy gaps satisfy proper rationally numeric ratios.
    \item The $d$ modular momentum states are in fact a generalization of the three current states defined in Eqs. \ref{current_states_triple} for the $d=3$ case, and are hence eigenvectors of the cyclic current operator in Eq. \ref{cyclic_current_operator}, straightforwardly extended to $d\times d$ matrices.
\end{itemize}

\subsubsection{Two- and multi- qudit gates}

As mentioned above, multi-qudit gates are necessary in order to execute general quantum algorithms, when two-qudit gates are the simplest choice.

Methods of magnetically coupling two flux qubits have been shown using mutual inductive coupling or coupling with transformer loops, and two-qubit operations have been demonstrated such as controlled-Z and controlled-NOT \cite{controlled not, controlled z and not}. Furthermore, execution of Shor's factoring algorithm have been demonstrated using five flux qubits with bi- and tripartite entanglements, factoring the number 15 with 48\% success \cite{Shor factoring 15}. Quantum Fast Fourier Transform using multilevel atoms was also suggested \cite{QFFTMLA}.

The broad subject of flux qudit couplings produced gates is out of scope for this work. For further details on the entangling process and construction of the gates, one is referred to \cite{Qutent,Qudent}.

\subsection{Scalability and the optimal range of $d$'s}


We have discussed above the computational advantage of a higher $d$ qudit. What then is the downside of higher $d$, and what range of $d$ values would be optimal?
First, as we increase $d$ we may have to increase the total length $L$ of the device, which might increase the duration of time needed for completing a computation, and moreover, may be longer than the achievable decoherence times.  Second, higher $d$ values make the energy differences between the levels smaller, as can be seen from Eq. \ref{energiesEn}. It might hence challenge our basic assumption in Eq. \ref{ddifferences}. These considerations may indeed suggest that there is a non-trivial $d$ value for which the system of wells is optimal, given a specific implementation and a computation to perform.

\subsection{Notes on implementation}

In the double- and triple-well sections we have referred to the main implementation using superconducting devices, which form the controllable potential well forms about the flux quantum variable. However, a similar implementation for a superconducting $d$-well system has yet to be proposed.
Nonetheless, an implementation of interacting $d$-well qudits using quantum dots have been demonstrated by Schirmer et al. \cite{d-well dots}. Their proposed solution features a charged particle in a quantum dot array forming a potential well chain . To agree with the above qudit analysis, however, the fabrication of the quantum dot should be modified to an enclosed circular form. In addition, to reach a fault tolerant operation in a real setup, a scheme for error correction, e.g. the one implemented in \cite{ErrCorrSC} would probably be needed.









\section*{Summary and Discussion}
\addcontentsline{toc}{chapter}{Summary and Discussion}

The use of double-well systems for computation, particularly via superconductor systems, has been an active research area during the last years. Its characteristics were studied via multiple experiments, a selection of which were mentioned throughout the work \cite{Josephson qubits, Ultrafast SQUID, Wendin, SCqubits_Review, Flux-based qubit, IBM qubit 2006, Coherence QED cavity 2011}. While many studies of certain operations and gates on the flux qubit were performed, the feasibility of arbitrary operations in the context of high computational efficiency has not been much discussed to our knowledge. Furthermore, it appears that the possibly advantageous extension of the ideas towards large $d$ system with periodic boundary conditions has not been thoroughly considered.

This work has examined the foundations of the potential well system in the two-state regime: from the evolution of its `spin' operators and the corresponding physical observables, to its manipulation within the common control scheme of SQUID systems. It was shown how these changes of the potential's shape translate to qubit operations, and eventually, how they may be utilized to carry out arbitrary $SU(2)$ operations in minimal steps for efficient usage of the coherence time. An extension of the SFQ method \cite{pulse train} used for RF-pulsing ($Z$-rotation) was proposed for rapid execution of Hamiltonian axis tilting (rotation about an arbitrary axis in $XZ$). We proceeded to propose an analogous setup for realisation of a qutrit using a controllable cyclic triple-well potential. Analysis of the three-state dynamics was shown, and schemes for execution of a subset of $SU(3)$ operators, among which the ternary X-gates \cite{Khan, Qutrit computing} and the 3-state quantum Fourier transform. We eventually generalized the ideas to $d$ well systems in two proposed directions - the fully connected wells and the periodic well chain - noting their characteristics and higher dimensional dynamics in its $SU(d)$ context.

The results of our work may hopefully strengthen the motivation for using qutrits or qudits as alternatives for the traditional qubit, in the context of potential-wells realisation. Achieving a fully controllable $d$-well system - either based upon flux qubits in superconducting circuitry, charge qubits via quantum dots, or another suitable mechanism - may indeed prove as a significant step towards the goal of a general-purpose quantum computation system. Further research of implementation methods is encouraged, with the open questions of optimal $d$ and system scalability to be tackled within the scope of each considered method.\\

\noindent{{\bf Acknowledgments}} \hfill\break
We thank Boaz Tamir for many helpful discussions. Y.A. acknowledges support from the Israel Science Foundation (grant no. 1311/14), the ICORE Excellence Center `Circle of Light' and the German-Israeli Project Cooperation (DIP) for support. E.C. was supported by ERC AdG NLST.

\section*{Appendix A: The Square Double-Well Solution}

\begin{figure}[h]
 \centering \includegraphics[height=7cm]{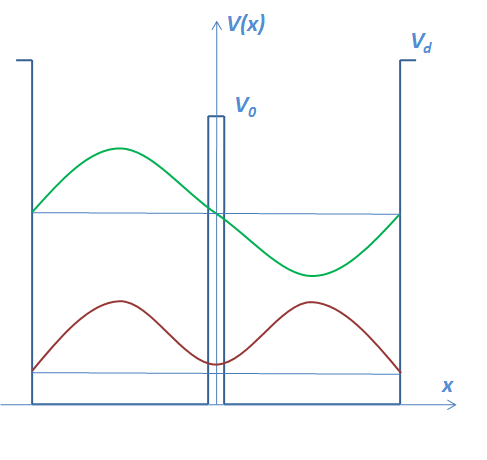}
      \caption{{\bf A basic example - square double-well potential. The two lowest energy states are illustrated.}} \label{Double_Well_States}
\end{figure}

We analyze the example of an ideal one-dimensional square potential double-well. In the 1D Hamiltonian $H = \frac{p^2}{2m} + V(x)$ we have for $V(x)$:
\begin{equation} \label{Square_Well_eqs}
V(x) = \begin{cases}
V_D & (L+\frac{a}{2}) \leq x < \infty \\
0 & \frac{a}{2} < x <  (L+\frac{a}{2}) \\
V_0 & -\frac{a}{2} \leq x \leq \frac{a}{2} \\
0 & -(L+\frac{a}{2}) < x <  -\frac{a}{2} \\
V_D & - \infty < x \leq -(L+\frac{a}{2}),
\end{cases}
\end{equation}

as shown in Fig. \ref{Double_Well_States}, marking the two wells by (I, III) and the barrier by (II).

Our regime of interest refers to a confined particle at energies less (particularly much less) than $V_0$. Namely,
\begin{equation} \label{Potential_relations}
V_D \gg  V_0 > E.
\end{equation}
The wavefunction $\Psi(x)$ thus splits into three parts, i.e. $\Psi_L(x)$ in the left well (region I), $\Psi_R(x)$ in the right well (region III) and $\Psi_B(x)$ for the barrier (region II). Note that based on Eq. \ref{Potential_relations} we may take the amplitude in the outer regions to be negligible: $\Psi(x) \approx 0$ for $|x| > L+\frac{a}{2}$.

The energy spectrum consists of pairs of states, $\ket{2k}$ and $\ket{2k+1}$, nearly degenerate with respect to the energy difference between pairs. Our interest is focused on the lowest states $\ket{0}$, $\ket{1}$ (illustrated in Fig. \ref{Double_Well_States}), for which the energy difference $\Delta E_{01} = E_1 - E_0$ satisfies Eq. \ref{Energies_low_states}.

For the full analysis of the square double-well, one is referred to \cite{exact_double_well_schrodinger}.


\end{document}